\documentclass[a4paper,fleqn,usenatbib]{mnras}

\usepackage{mathptmx}

\usepackage[T1]{fontenc}
\usepackage{ae,aecompl}

\usepackage{graphicx}	
\usepackage{amsmath}	
\usepackage{amssymb}	

\title[The orbit of HD\,149277]{HD\,149277: a rare short-period SB2 system with
a subsynchronously rotating magnetic He-rich primary\thanks{Based on observations obtained at the Canada-France-Hawaii Telescope (CFHT) which is operated by the National Research Council of Canada, the Institut National des Sciences de l'Univers of the Centre National de la Recherche Scientique of France, and the University of Hawaii.}.}

\author[J. F. Gonz\'alez et al.]{
J. F. Gonz\'alez,$^{1}$\thanks{E-mail: jfgonzalez@conicet.gov.ar}
S. Hubrig,$^{2}$\thanks{E-mail: shubrig@aip.de}
S. P. J\"arvinen$^{2}$,
M. Sch\"oller$^{3}$
\\
$^{1}$Instituto de Ciencias Astron\'omicas, de la Tierra y del Espacio (ICATE), Av. Espa\~na Sur 1512, CC 49, 
5400 San Juan, Argentina\\
$^{2}$Leibniz-Institut f\"ur Astrophysik, An der Sternwarte 16, 14482 Potsdam, Germany\\
$^{3}$European Southern Observatory, Karl-Schwarzschild-Str.~2, 85748 Garching, Germany
}

\date{Accepted XXX. Received YYY; in original form ZZZ}

\pubyear{2015}

\begin{document}
\label{firstpage}
\pagerange{\pageref{firstpage}--\pageref{lastpage}}
\maketitle

\begin{abstract}
HD\,149277 is a rare SB2 system with a slowly rotating magnetic He-rich primary with $P_{\rm rot}=25.4$\,d. The CFHT/ESPaDOnS archive
spectra revealed $P_{\rm orb}=11.5192 \pm 0.0005$\,d indicating strong subsynchronous rotation of the primary component. Such a strong
subsynchronous rotation was not detected in any other SB2 system with a magnetic chemically peculiar component.
Our inspection of the spectra revealed the presence of resolved Zeeman split spectral lines allowing us to determine the
variability of the mean magnetic field modulus over the rotation period. The maximum of the magnetic field modulus concides roughly with
the positive extremum of the longitudinal field, whereas the
minimum of the modulus with the negative extremum of the longitudinal field. 
No evidence for a longitudinal magnetic field was seen in the circularly polarized spectra of the secondary component.  Using archival data
from the ASAS3 survey,
we find in the frequency spectrum only one significant peak,
corresponding to the period  $P_{\rm phot}=25.390 \pm 0.014$\,d.
This value is in good agreements with the previous determination of the
rotation period, $P_{\rm rot}=25.380\pm0.007$\,d, which was based on longitudinal magnetic field measurements.
\end{abstract}





\section{Introduction}

Using  high-resolution polarimetric spectra of HD\,149277
acquired with ESPaDOnS at the Canada-France-Hawaii Telescope (CFHT), \citet{Shultz2018} recently reported 
on the presence of a rather strong longitudinal magnetic field with $\left<B_{\rm z}\right>_{\rm max}=3.3\pm0.1$\,kG.
The authors determined a rotation period of 25.4\,d using these mean longitudinal magnetic field measurements.
HD\,149277 is a member of the young open cluster NGC\,6178 at an age of $\log\,(t [\mathrm{yr}])=7.15\pm0.22$ \citep{Kharchenko2005}.
Its physical properties, $T_{\rm eff}=22\,300\pm 500$\,K, $M=8.75\pm0.40$\,M$_{\odot}$, and 
$\log (L/L_{\odot})=3.7\pm 0.1$  were reported by \citet{Landstreet2007}.
In the column ``remarks'' of their Table~1, \citet{Shultz2018}
mention that HD\,149277 is an SB2 system
with a B2IV/V magnetic primary, but give no information on the nature of the secondary component or the orbital parameters.

According to previous studies by e.g.\ \citet{Carrier2002}, \citet{Hubrig2014}, and \citet{Landstreet2017}, close SB2 systems with
magnetic Ap or Bp components are extremely rare. Among the studied binary systems with A and late B-type primaries,
only two systems, HD\,98088 and HD\,161701, are known to possess
a magnetic Ap star as a companion \citep{Babcock1958,Abt1968,Hubrig2014}, and only a few more SB2 systems 
with Bp components are currently known \citep{Landstreet2017}.
Such systems are of considerable interest in the context of the magnetic field origin,
which is still not properly understood. \citet{Ferrario2009} suggested that
magnetic Ap and Bp stars are products of a merger of two lower mass protostars.
In this scenario, the binaries that we observe now were triple systems earlier
in their history. In the following we report on our analysis of the orbital parameters of HD\,149277 and give
additional information on the spectral, magnetic, and photometric variability.

\section{Orbital analysis}

\begin{table}
\centering
\caption{
Radial velocity measurements carried out for the primary and secondary components in the system HD\,149277. The first observation corresponds to the HARPS spectrum.
}
\label{tab:RV}
\begin{tabular}{ccr@{$\pm$}lr@{$\pm$}l}
\hline
\multicolumn{1}{c}{HJD} &
\multicolumn{1}{c}{Orbital} &
\multicolumn{2}{c}{$RV_{\rm A}$} &
\multicolumn{2}{c}{$RV_{\rm B}$} \\
2\,450\,000+&
\multicolumn{1}{c}{Phase} &
\multicolumn{2}{c}{(km\,s$^{-1}$)} &
\multicolumn{2}{c}{(km\,s$^{-1}$)} \\
\hline
4206.3982 &  0.636 &     69.3 &    0.3 &  $-$79.7 &    0.5\\
6112.3310 &  0.103 & $-$113.0 &    0.3 &    119.4 &    0.4\\
6759.6146 &  0.298 &  $-$75.4 &    0.3 &     76.1 &    0.5\\
6760.4658 &  0.372 &  $-$36.9 &    0.3 &     33.6 &    0.6\\
6812.3925 &  0.880 &     71.6 &    0.3 &  $-$83.0 &    0.5\\
6818.3859 &  0.400 &  $-$21.6 &    0.3 &     18.7 &    0.4\\
6821.3857 &  0.660 &     74.7 &    0.3 &  $-$87.5 &    0.5\\
6822.3910 &  0.748 &     88.7 &    0.3 & $-$103.2 &    0.7\\
6824.3772 &  0.920 &     51.1 &    0.3 &  $-$61.8 &    0.5\\
7225.2883 &  0.726 &     87.0 &    0.3 & $-$100.4 &    0.5\\
7226.2343 &  0.808 &     88.8 &    0.3 & $-$103.7 &    0.5\\
7227.2542 &  0.896 &     64.4 &    0.3 &  $-$77.4 &    0.5\\
7228.2936 &  0.987 &   $-$4.1 &    0.4 &   $-$1.0 &    0.9\\
7229.3121 &  0.075 &  $-$93.2 &    0.3 &     98.0 &    0.5\\
7230.3155 &  0.162 & $-$125.4 &    0.4 &    134.4 &    0.8\\
7231.2850 &  0.246 & $-$100.0 &    0.4 &    104.8 &    0.5\\
7232.3179 &  0.336 &  $-$54.0 &    0.4 &     55.4 &    0.5\\
7233.2322 &  0.415 &  $-$14.0 &    0.4 &      9.1 &    0.5\\
7234.2364 &  0.503 &     24.4 &    0.4 &  $-$32.6 &    0.7\\
7235.2543 &  0.591 &     55.8 &    0.4 &  $-$66.5 &    0.5\\
7236.3116 &  0.683 &     79.5 &    0.3 &  $-$93.3 &    0.5\\
7604.2298 &  0.624 &     66.6 &    0.3 &  $-$78.3 &    0.5\\
7611.2236 &  0.231 & $-$106.5 &    0.4 &    112.6 &    0.5\\
7611.2503 &  0.234 & $-$105.4 &    0.4 &    111.4 &    0.5\\
\hline
\end{tabular}
\end{table}

\begin{figure}
\centering
\includegraphics[width=0.9\columnwidth,height=6.4cm, bb=30 2 510 737]{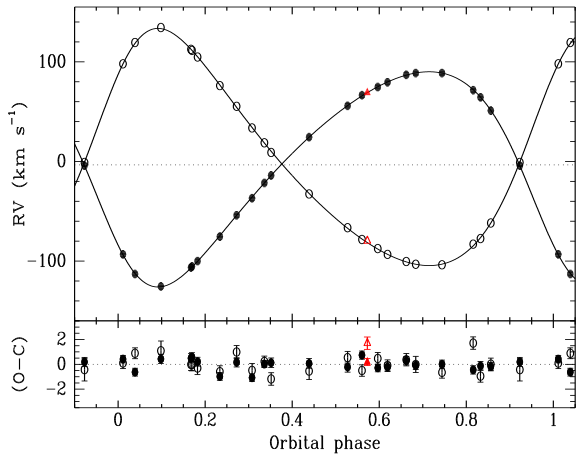}
\caption{
Spectroscopic orbit of the system HD\,149277 using ESPaDOnS spectra obtained 
from 2012 to 2016 and one HARPS spectrum (red triangles).
Open symbols indicate measurements of the secondary component,
which is the less massive star.
The lower panel shows the residuals: observed minus calculated. Phase zero corresponds to the
periastron passage. In the upper panel error bars are smaller than the symbol sizes.
}
\label{fig:RV}
\end{figure}

To estimate the orbital parameters of the SB2 system HD\,149277, we downloaded all publically available  
ESPaDOnS (Echelle SpectroPolarimetric Device for the Observation of Stars) spectra obtained from 2012 to 2016.
ESPaDOnS is a spectropolarimeter with a resolving power of about 65\,000, covering the wavelength region
from 3700 to 10\,500\,\AA{}.
As most of the spectra have a rather low signal-to-noise ratio (S/N)
between 85 and 100,
we averaged spectra recorded during the same night (separated by up to 3\,h) in order to improve the S/N,
obtaining a total of 23 spectra.
To this sample we added one available HARPS (High Accuracy Radial velocity Planet Searcher)
spectrum taken in 2007.
In Table~\ref{tab:RV} we present our radial velocity measurements for the primary and secondary components on 24 phases and
in Fig.~\ref{fig:RV} the spectroscopic orbit of the system.
The value $v\,\sin\,i_{\rm B}$ consigned in Table~\ref{tab:RV} was determined using the S\,{\sc ii}~5665 line with a low Land\'e factor of 0.5.

\begin{table}
\centering
\caption{
Orbital and fundamental parameters for the SB2 system HD\,149277.
}
\label{tab:orbit149277}
\begin{tabular}{cr@{$\pm$}lc}
\hline
$P$ (d)                & 11.5192 & 0.0005 \\
$T_{\rm conj}$             & 2456802.77 & 0.02 \\
$T_{\rm per}$              & 2456803.51 & 0.06 \\
$V_{\rm o}$ (km\,s$^{-1}$)   & $-$2.6 & 0.3 \\
$K_A$ (km\,s$^{-1}$)         & 107.8 & 0.7 \\
$K_B$ (km\,s$^{-1}$)         & 118.2 & 0.8 \\
$\omega$                     & 2.23& 0.03 \\
$e$                          & 0.236 & 0.004 \\
$a\,\sin\,i$ ($R_\odot$)     & 50.0 & 0.2 \\
$M\,\sin^3\,i$ ($M_\odot$)   & 12.66 & 0.16 \\
$M_{\rm A}\,\sin^3\,i$ ($M_\odot$)  & 6.62 & 0.09 \\
$M_{\rm B}\,\sin^3\,i$ ($M_\odot$)  & 6.04 & 0.08 \\
$q$                          & 0.912 & 0.007 \\
\hline
$v\,\sin\,i_{\rm A}$ (km\,s$^{-1}$)  & 8           &     3$^a$\\
$v\,\sin\,i_{\rm B}$ (km\,s$^{-1}$)  & 11.5          &   4\\
\hline
\end{tabular}\\

$^a$ \citet{Shultz2018}.
\end{table}

 Assuming the photometric mass derived by  \citet{Landstreet2007}, which is consistent with the spectral type,
the orbital inclination is estimated to be around $66^{\circ}$.

The derived orbital parameters (see Table~\ref{tab:orbit149277})
reveal that the orbit of HD\,149277 is eccentric and that the rotation
of the components is not synchronized.
In the case of pseudosynchronization we would expect 
$P_{\rm pseudosync}=8.92$\,d.
The primary component shows strong subsynchronous rotation, most probably
caused by magnetic breaking in a system without strong rotation-orbital coupling. Given the mass ratio
of 0.912, both components are early B-type stars. The secondary is probably by 1000\,K cooler with a spectral type B3$^-$V.  

\section{Magnetic field measurements}
\label{sec:MF} 

\begin{figure}
\centering
\includegraphics[width=\columnwidth, height = 12cm, bb=120 30 495 737]{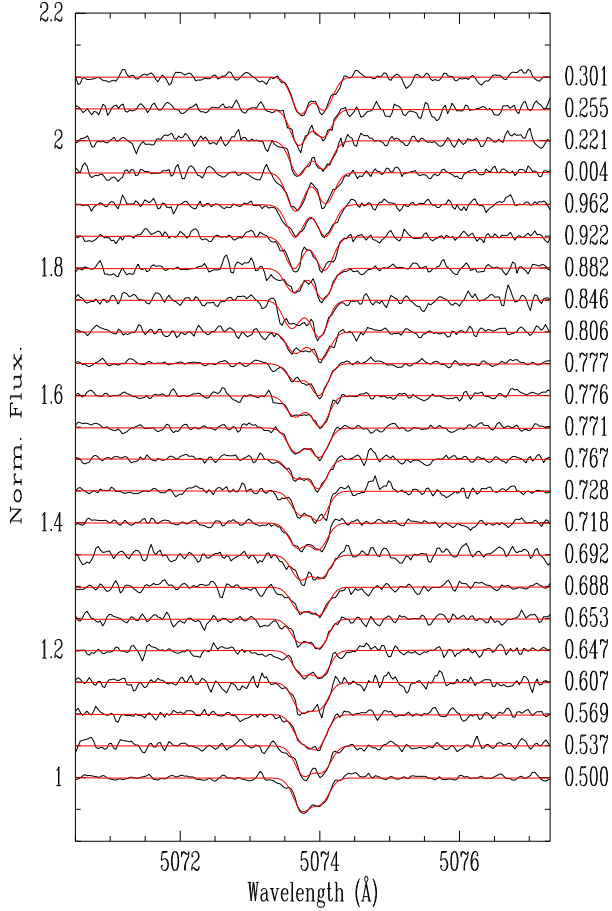}
\caption{
Variability of the magnetically split Fe\,{\sc iii}~5073.6 line over the rotation period in the disentangled 
spectra of the primary component.
Spectra are vertically offset for better visibility and are sorted in rotational phase
(see right side; starting at 0.5) from bottom to top.
The red lines represent a double-Gaussian fit to the data.
}
\label{fig:res}
\end{figure}

One of the most interesting aspects of the system HD\,149277 is the rather strong magnetic field and slow rotation of its
primary component. Our inspection of the ESPaDOnS spectra revealed the presence of several resolved Zeeman split lines.
Their presence has not been reported previously.
A number of Ap  and late-type Bp stars with strong magnetic fields show resolved Zeeman split spectral lines,
allowing to set additional constraints on the magnetic field geometry by measuring the mean magnetic 
field modulus, i.e.\ the average over the  visible stellar hemisphere of the modulus of the magnetic field 
vector, weighted by the local line intensity, following the relation given
e.g.\ by \citet{HubrigNesvacil2007} and \citet{Mathys2017}.
Such Zeeman split lines can only be seen in stars whose projected
rotational velocity is sufficiently small and whose magnetic field
is strong enough to exceed the rotational Doppler broadening \citep{Mathys1997}.
In Fig.~\ref{fig:res} we present 
the rotational variability of the resolved Zeeman split Fe\,{\sc iii}~5073.6 line in the primary component.
Interestingly, as illustrated in this figure and below in Fig.~\ref{fig:4553}, we observe that in certain phases the Zeeman split
lines are asymmetric about the line centers with the blue
components appearing less deep than the red components, and with the red components deeper in other phases.
Such asymmetries occur in the same way in other transitions, indicating that the rotational Doppler effect is non--negligible, i.e.
various parts of the stellar surface characterized by different magnetic field strengths contribute to the
line profile.

\begin{table}
\centering
\caption{
Measurements of the mean magnetic field modulus in the primary component of HD\,149277.
}
\label{tab:MF}
\begin{tabular}{ccr@{$\pm$}lr@{$\pm$}l}
\hline
\multicolumn{1}{c}{HJD} &
\multicolumn{1}{c}{Rotational} &
\multicolumn{2}{c}{$\left\langle B\right\rangle$} \\
2\,450\,000+&
\multicolumn{1}{c}{Phase} &
\multicolumn{2}{c}{(kG)} \\
\hline
6112.3310 &  0.718  &   6.32 &  0.24 \\
6759.6146 &  0.221  &   7.48 &  0.28 \\
6760.4658 &  0.255  &   7.27 &  0.30 \\
6812.3925 &  0.301  &   7.01 &  0.24 \\
6818.3859 &  0.537  &   5.79 &  0.34 \\
6821.3857 &  0.655  &   6.08 &  0.36 \\
6822.3910 &  0.695  &   6.23 &  0.49 \\
6824.3772 &  0.773  &   6.96 &  0.22 \\
7225.2883 &  0.569  &   5.00 &  0.37 \\
7226.2343 &  0.607  &   5.51 &  0.39 \\
7227.2542 &  0.647  &   5.90 &  0.37 \\
7228.2936 &  0.688  &   6.01 &  0.41 \\
7229.3121 &  0.728  &   5.92 &  0.36 \\
7230.3155 &  0.767  &   6.55 &  0.36 \\
7231.2850 &  0.806  &   7.60 &  0.32 \\
7232.3179 &  0.846  &   8.11 &  0.34 \\
7233.2322 &  0.882  &   8.80 &  0.31 \\
7234.2364 &  0.922  &   9.03 &  0.24 \\
7235.2543 &  0.962  &   9.04 &  0.22 \\
7236.3117 &  0.004  &   8.94 &  0.22 \\
7604.2298 &  0.500  &   5.64 &  0.23 \\
7611.2236 &  0.776  &   7.46 &  0.26 \\
7611.2503 &  0.777  &   6.96 &  0.28 \\
\hline
\end{tabular}
\end{table}

\begin{figure}
\centering
\includegraphics[width=0.95\columnwidth,height=5.5cm, bb=20 30 557 737]{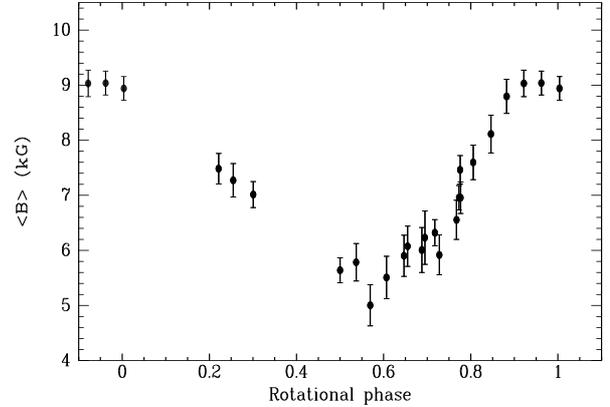}
\caption{
Mean magnetic field modulus, calculated from the Zeeman split Fe\,{\sc iii}~5073.6 line,
over the rotation of the primary component of HD\,149277.
}
\label{fig:MF}
\end{figure}

In order to estimate the mean magnetic field modulus, we fitted the line $\lambda$5073.9
with two Gaussian profiles. The position and height of both Zeeman components were considered
as free parameters, but the Gaussian width was assumed to be the same for both components in all spectra. 
The use of a fixed value of Gaussian width gives better results in the spectra with lower S/N and eases to
measure the position of the components even if they are blended.
The values of the mean magnetic field modulus
(see Table~\ref{tab:MF} and Fig.~\ref{fig:MF})
were calculated adopting a Land\'e factor $g_\mathrm{eff}=2.01$, taken from the
VALD database \citep{Kupka2011},
and are comprised between 5.0 and 9.0\,kG.
These measurements were cross-checked
with measurements of the Zeeman split Fe\,{\sc ii}~5018.4 line, 
which however affected by low S/N and blending with the spectral lines of the secondary 
in several rotation phases.

The maximum of the mean magnetic field modulus coincides roughly with the positive extremum of the longitudinal magnetic field, 
whereas the minimum of the modulus coincides with the negative extremum of the longitudinal magnetic field.
The phase curve for the longitudinal magnetic field
is presented  in Fig.~A20 in the work of \citet{Shultz2018}.
Future spectroscopic monitoring at higher S/N will be worthwhile to better characterize the variability of the 
mean magnetic field modulus in this component.
No evidence for a longitudinal magnetic field was seen in the circularly polarized spectra of the secondary component.

\section{Spectral and photometric variability}

\begin{figure}
\centering
\includegraphics[width=\columnwidth,height=10cm]{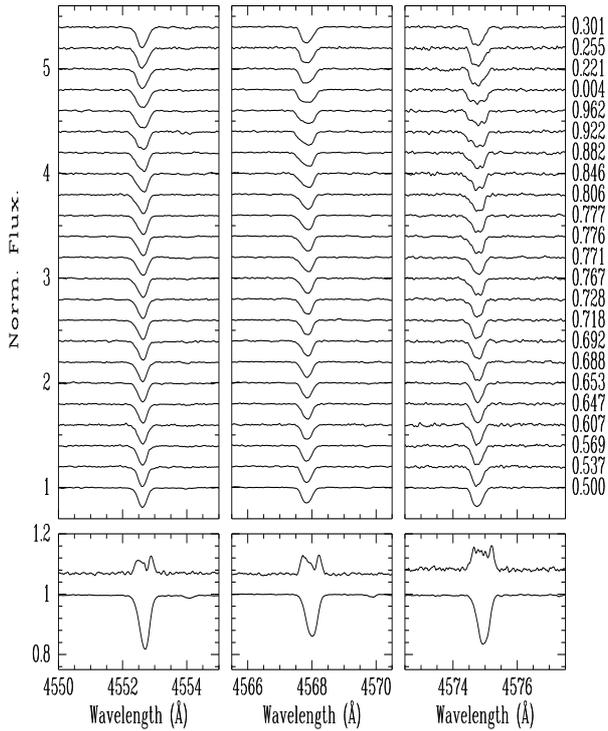}
\caption{
Variability of three \ion{Si}{iii} line profiles in the disentangled spectra of 
the primary component of HD\,149277.
Labels on the right refer to the rotational phases calculated
with $P_{\rm rot}$ and  $JD0$ given by \citet{Shultz2018}.
The spectra have been shifted vertically for better visibility.
The spectra of the faintest line $\lambda$4575 have been scaled by a factor 2.
The lower panels show the mean profile and the RMS of residuals (scaled $\times$5)
}
\label{fig:4553}
\end{figure}

Like other magnetic Bp stars, HD\,149277 exhibits distinct changes 
in the line profiles of different elements, probably caused by an inhomogeneous element distribution
on the stellar surface.
As an example, nitrogen lines become stronger at the phases conciding with the negative extremum of the
longitudinal magnetic field.
The clearest variations are observed in the profiles of the \ion{Si}{iii} lines.
The behaviour of three \ion{Si}{iii} line profiles in the disentangled spectra of the primary is presented in Fig.~\ref{fig:4553}.

 
\begin{figure}
\centering
\includegraphics[width=0.95\columnwidth,height=10cm,angle=0]{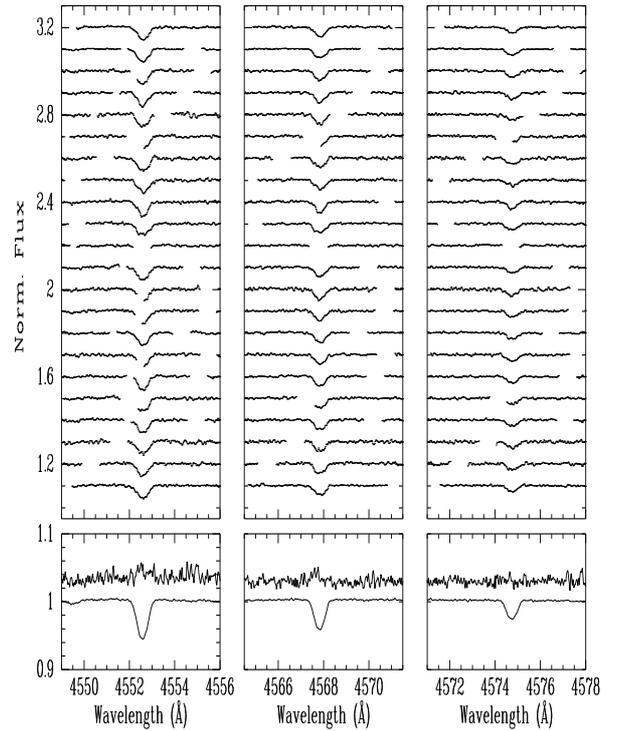}
\caption{Line profile of \ion{Si}{iii}~4553  in the disentangled spectra of 
the secondary component of HD\,149277. 
The spectra are ordered by JD from bottom to top.
The spectra of the faintest line $\lambda$4575 have been scaled by a factor 2.
The lower panels show the mean profile and the RMS of residuals (scaled $\times$10)
}
\label{fig:sec}
\end{figure}

Since spectral lines of the secondary component are weaker and in many cases are overlapped 
with variable lines of the primary, the existence of spectral variability
cannot definitely be established.
In Fig.~\ref{fig:sec} we present the profiles of the \ion{Si}{iii}~4553 line recorded in the spectra of the secondary
component on different orbital phases.
Missing parts in the spectra correspond to the position of spectral lines of the primary star (all lines deeper than
1\% of the continuum level).
Since lines of the primary are certainly variable these regions were not considered in the calculation of the RMS
(lower panel). 
The rotation period of this component could probably be determined from observed line profile variations,
provided higher S/N spectra of this system are recorded in future observations.
An analysis of the spectral variability of the secondary is of
special interest: for the majority of the previously studied close binary systems with $P_{\rm orb} \le 20$\,d possessing
one magnetic or chemically peculiar spectrum variable component,
the other component is usually an Am star or a normal A or B-type star.
To our knowledge, there are only three exceptions: the system 41\,Eri consisting of two spectrum variable HgMn
stars \citep{Hubrig2012},
the system HD\,161701 consisting of a spectrum variable HgMn star and
a magnetic Ap star \citep{Hubrig2014}, and the system HD\,136504 consisting of two magnetic Bp stars.
The magnetic field in the primary of HD\,136504 was detected in low-resolution FORS\,2 observations by \citet{Hubrig2010},
while the discovery of a magnetic field in the secondary was announced by \citet{Shultz2015}.


\begin{figure}
\centering
\includegraphics[width=0.9\columnwidth,height=5.2cm]{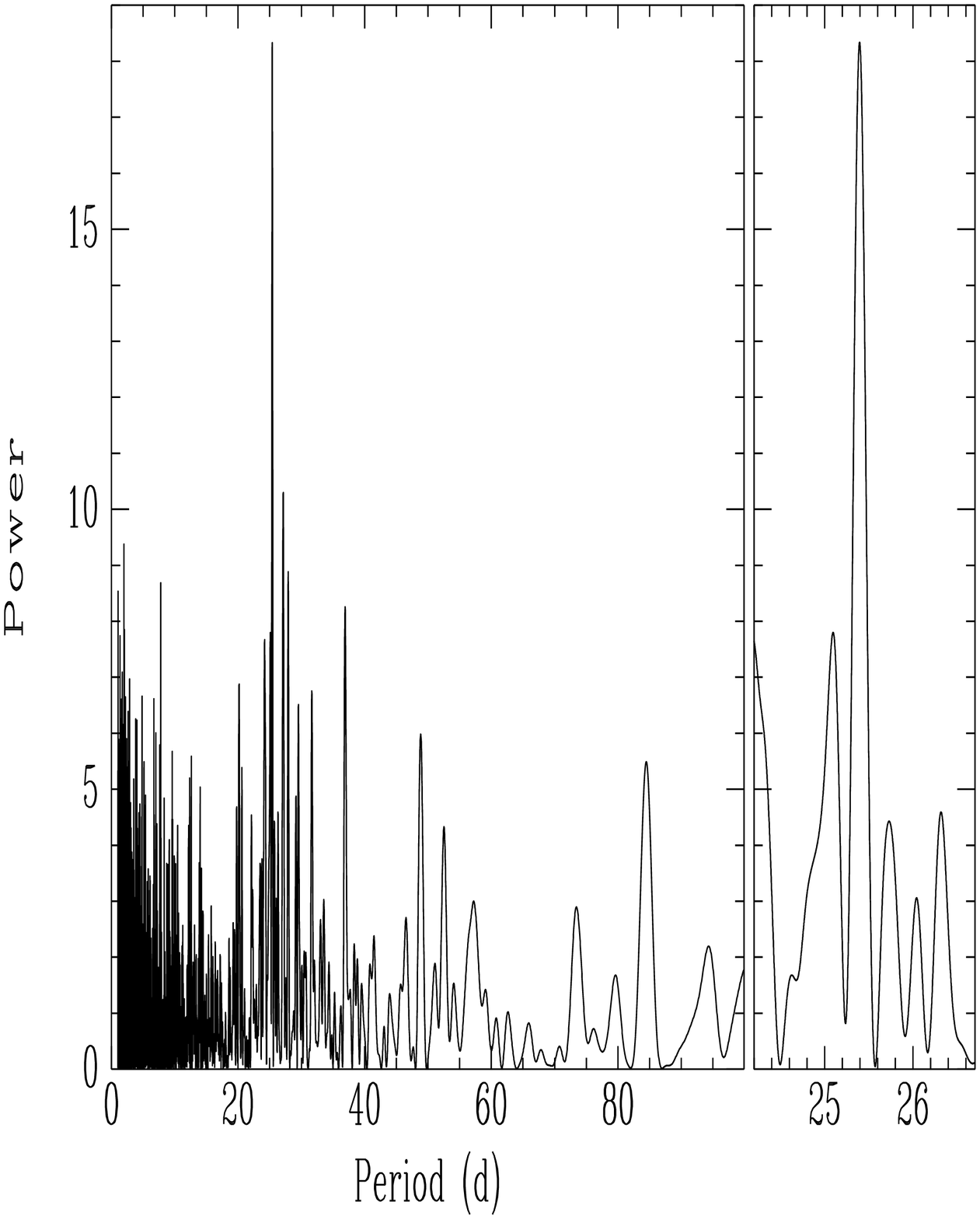}
\includegraphics[width=0.9\columnwidth,height=5.2cm]{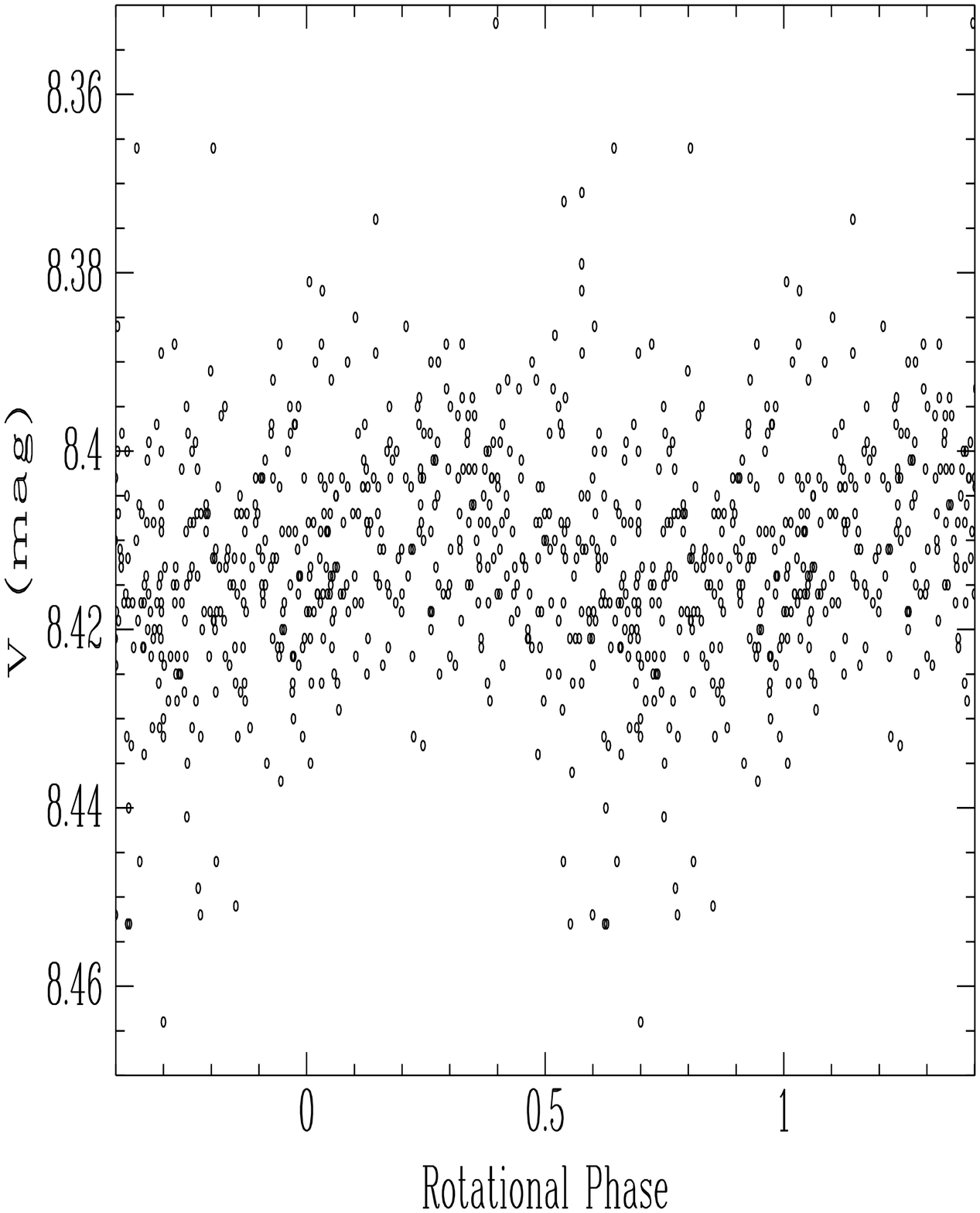}	
\caption{
Photometric variability of the primary. 
Upper panel: Periodogram obtained using the ASAS3 database. 
Lower panel: The ASAS3 light curve using $P_{\rm phot}=25.390$\,d.
}
\label{fig:phot_var}
\end{figure}

He-rich stars and chemically peculiar stars usually exhibit photometric variability
due to the presence of surface chemical abundance spots. Therefore, rotation periods can also 
be determined from photometry.
Using archival data from the ASAS3
survey\footnote{http://www.astrouw.edu.pl/asas/}
\citep{Pojmanski1997},
we find in the frequency spectrum only one significant peak,
$P_{\rm phot}=25.390 \pm 0.014$\,d (see Fig.~\ref{fig:phot_var}).
This value is in good agreements with the previous determination of the
rotation period, $P_{\rm rot}=25.380\pm0.007$\,d
by \citet{Shultz2018}, which was based on longitudinal magnetic field measurements.
The photometric variability of this object with a light curve amplitude of 0.010$\pm$0.005\,mag is reported here
for the first time.

\section{Discussion}

The analysis of the system HD\,149277 is of considerable interest as close main sequence SB2 systems only very rarely
contain a magnetic Ap or
Bp star as a component. We determined the orbital parameters of the system and studied a few characteristics
of the individual components, such as spectroscopic and photometric variability. Similar to a few other SB2 systems with magnetic
Bp components compiled in Table~5 in the work of \citet{Landstreet2017}, the orbit is quite eccentric with $e=0.237$,
but the rotation period of the primary is the longest among the studied systems.
The more massive of
the two stars is rotating subsynchronously and exhibits a strongly variable magnetic field modulus.
Among the previously studied SB2 systems, the primary in HD\,149277 exhibits the strongest magnetic field and
the strongest subsynchronous rotation.
The less massive component is apparently non-magnetic,
but probably shows spectral variability, which should be analysed in more detail
in future higher S/N observations.
Such future high S/N observations should also be used to constrain the structure of
the magnetic field of the primary in more detail.

We note that studies of strongly magnetic stars using the mean magnetic field modulus, measured using
magnetically resolved lines, are of importance because they provide the best opportunity
to study the effect of a magnetic field on stellar atmospheres. 
At present, 84~Ap stars are known to show magnetically resolved lines,
representing 2.3\% of the total number of known Ap stars \citep{Mathys2017}.
The sample of early-type massive Bp stars
currently consists of just three members (Hubrig et al., {\sl in preparation}).


\section*{Acknowledgements}
This work has made use of the VALD database, operated at Uppsala
University, the Institute of Astronomy RAS in Moscow, and the University of
Vienna.
Based on observations made with ESO Telescopes at the La Silla Paranal Observatory
under programme ID 079.D-0118(A).
JFG thanks financial support from CONICET (PIP 0331) and the Universidad Nacional de San Juan (Proyecto CICITCA), Argentina





\begin{thebibliography}{99}

\bibitem[\protect\citeauthoryear{Abt et al.}{1968}]{Abt1968}
Abt H.~A., Conti P.~S., Deutsch A.~J., Wallerstein G.,
1968, ApJ, 153, 177

\bibitem[\protect\citeauthoryear{Babcock}{1958}]{Babcock1958}
Babcock H.~W.,
1958, ApJS, 3, 141

\bibitem[\protect\citeauthoryear{Carrier et al.}{2002}]{Carrier2002}
Carrier F., North P., Udry S., Babel J.,
2002, A\&A, 394, 151

\bibitem[\protect\citeauthoryear{Ferrario et al.}{2009}]{Ferrario2009}
Ferrario L., Pringle J.~E., Tout C.~A., Wickramasinghe D.~T.,
2009, MNRAS, 400, L71

\bibitem[\protect\citeauthoryear{Hubrig \& Nesvacil}{2007}]{HubrigNesvacil2007}
Hubrig S., Nesvacil N.,
2007, MNRAS, 378, L16

\bibitem[\protect\citeauthoryear{Hubrig et al.}{2010}]{Hubrig2010}
Hubrig S., Ilyin I., Sch\"oller M., Briquet M., Morel T., De Cat P.,
2010, ApJ, 726, L5

\bibitem[\protect\citeauthoryear{Hubrig et al.}{2012}]{Hubrig2012}
Hubrig S., et al.,
2012, A\&A, 547, A90

\bibitem[\protect\citeauthoryear{Hubrig et al.}{2014}]{Hubrig2014}
Hubrig S., et al.,
2014, MNRAS, 440, L6

\bibitem[\protect\citeauthoryear{Kharchenko et al.}{2005}]{Kharchenko2005}
Kharchenko N.~V., Piskunov A.~E., R\"oser S., Schilbach E., Scholz R.-D.,
2005, A\&A, 438, 1163

\bibitem[\protect\citeauthoryear{Kupka et al.}{2011}]{Kupka2011}
Kupka F., Dubernet M.-L., VAMDC Collaboration,
2011, Baltic Astronomy, 20, 503

\bibitem[\protect\citeauthoryear{Landstreet et al.}{2007}]{Landstreet2007}
Landstreet J.~D., Bagnulo S., Andretta V., Fossati L., Mason E., Silaj J., Wade G.~A.,
2007, A\&A, 470, 685

\bibitem[\protect\citeauthoryear{Landstreet et al.}{2017}]{Landstreet2017}
Landstreet J.~D., Kochukhov O., Alecian E., Bailey J.~D., Mathis S., Neiner C., Wade G.~A., BINaMIcS Collaboration,
2017, A\&A, 601, A129

\bibitem[\protect\citeauthoryear{Mathys et al.}{1997}]{Mathys1997}
Mathys G., Hubrig S., Landstreet J. D., Lanz T., Manfroid J.,
1997, A\&AS, 123, 353

\bibitem[\protect\citeauthoryear{Mathys}{2017}]{Mathys2017}
Mathys G.,
2017, A\&A, 601, A14

\bibitem[\protect\citeauthoryear{Pojmanski}{1997}]{Pojmanski1997}
Pojmanski G.,
1997, AcA, 47, 467

\bibitem[\protect\citeauthoryear{Shultz et al.}{2015}]{Shultz2015}
Shultz M., Wade G.~A., Alecian E., BinaMIcS Collaboration,
2015, MNRAS, 454, L1

\bibitem[\protect\citeauthoryear{Shultz et al.}{2018}]{Shultz2018}
Shultz M., et al.,
2018, MNRAS, 475, 5144

\end{thebibliography}








\bsp	
\label{lastpage}
\end{document}